\documentstyle[12pt]{article}

\def\bea{\begin{eqnarray}}
\def\eea{\end{eqnarray}}
\def\ts{\textstyle}
\def\h{{\textstyle {1\over2}}}
\def\dphi{\partial\phi}
\def\d2phi{\partial^2\phi}
\def\emt{energy-momentum tensor}
\def\VX3{e^{ip_3\phi_3}}
\def\Vp{e^{i\vec p\cdot\vec\phi}}

\begin{document}

\pagestyle{empty}
\rightline{UG-1/94}
\rightline{hep-th/9401164}
\rightline{January 1994}
\vspace{2.0truecm}
\centerline{\bf BRST cohomology of the critical $W_4$ string}
\vspace{1.5truecm}
\centerline{H.J.~Boonstra}
\vspace{.5truecm}
\centerline{Institute for Theoretical Physics}
\centerline{Nijenborgh 4, 9747 AG Groningen}
\centerline{The Netherlands}
\vspace{1.5truecm}
\centerline{ABSTRACT}
\vspace{.5truecm}
We study the cohomology of the critical $W_4$ string using
the $W_4$ BRST charge in a special basis in which it contains three
separately nilpotent BRST charges. This allows us to obtain the
physical operators in three steps. In the first step we obtain the
cohomology associated to a spin-four constraint only, and it
contains operators of the $c={4\over5}$ $W_3$ minimal model.
In the next step, where the spin-three constraint is added, these
operators get dressed to operators of the $c={7\over10}$ Virasoro
minimal model. Finally, the Virasoro constraint is added to obtain
the cohomology of the critical $W_4$ string. We describe the
structure of the complete cohomology and compare with other
results.

\vfill\eject

\pagestyle{plain}

\section{Introduction}
In string theory the symmetry structure on the world-sheet plays an
important role. Gravity on the world-sheet may be viewed as a result of
gauging the conformal symmetry of a matter system.
A consistent quantization
then leads to an ordinary bosonic string. Likewise, one could start from a
conformal field theory with additional symmetries and gauge these to
obtain an extension of gravity on the world-sheet. This may give rise
to different string theories. The best known example is the
superstring, but it seems to be possible to construct string theories
based on more general extended conformal symmetries.

The $W_N$ algebras \cite{Zam,FL} are nonlinear
higher-spin extensions of the
Virasoro algebra with generators of spins $2,3,...N$. String theories
based on $W_N$ algebras were first suggested in \cite{BG}.
By now, the $W_3$ string has been quite thoroughly investigated,
see e.g. the reviews \cite{review} and references therein.
Its physical spectrum has been classified completely in
\cite{PopeC,BMP2}. One remarkable feature of the $W_3$ string is its
relation with the Ising model \cite{DDR,Rama,PopeWN}. This relation
is clarified by going to a new basis of fields introduced in
\cite{redef}. In this new basis, a special subsector
of the $W_3$ model was seen
to correspond to the Ising model \cite{Hull}, see also
\cite{Ising,redef,PopeH1,FWpara}.

It was also recognized that the relation with minimal models
generalizes to $W_n$ minimal models
in $W_N$-strings \cite{DDR,Rama,PopeWN}.
In general, for a $W_N$ string description in terms of $N-1$ scalar fields
$\phi_i,\ i=1,2,...N-1$ and $N-1$ ghost pairs $(c_k,b_k),\ k=2,3,...N$,
one for each spin-$k$ current, some simple
numerology leads to the following
picture. In the realisation obtained from the Miura transformation for
$W_N$ \cite{FL}, the generators of $W_N$ can be written in terms of
the generators of $W_{N-1}$ plus an
extra scalar field \cite{FL,DDR,PopeWN},
e.g. for the energy-momentum tensor this means that
\bea
T_N=T_{N-1}-\h (\partial\phi_{N-1})^2 + ix\sqrt{{\ts{(N-1)N\over2}}}
\partial^2 \phi_{N-1}\,,
\eea
where the parameter $x$ is fixed by criticality, i.e. by requiring
the central charge of the matter fields to cancel the total ghost
central charge. The ghost system
corresponding to the spin-$k$ generator contributes $-2(6k^2-6k+1)$
to this total ghost central charge.
Thus the background charges of the scalar fields have
fixed values in this realisation of the $W_N$ algebra,
\bea
\alpha_n={\ts{2N+1\over2}}\sqrt{{\ts{n(n+1)\over N(N+1)}}}\,.
\eea
It is then readily seen that the central charge of the
fields $\{\phi_{N-1};c_N,b_N\}$ is
\bea
\label{cNN}
c_N^N=1+12(\alpha_{N-1})^2-2(6N^2-6N+1)=
{\ts{2(N-2)\over N+1}}=c_{N-1,N}\,,
\eea
which is precisely the central charge of the first unitary $W_{N-1}$
minimal model. In general, the central charges of
unitary $W_N$ minimal models are given by
\bea
c_{N,q}=(N-1)(1-{\ts{N(N+1)\over q(q+1)}})\,,\ \ q>N\,.
\eea
Note that for $N=3$, the central charge (\ref{cNN}) is just that of the
Ising model.
More generally, the central charge of the fields
$\{\phi_{n-1},...\phi_{N-1};c_n,b_n,...c_N,b_N\}$, which may be
considered to correspond to the subsector of the $N-n+1$ highest
spin currents of the $W_N$ algebra \cite{own3}, adds up to
\bea
c_N^n&=&\sum_{k=n-1}^{N-1}(1+12(\alpha_k)^2)-2\sum_{k=n}^N(6k^2-6k+1)
\nonumber\\
&=&(n-2)(1-{\ts{n(n-1)\over N(N+1)}})=c_{n-1,N}\,,
\eea
which is the central charge
of the $(p,p')=(N,N+1)$ $W_{n-1}$ minimal model.
Thus we see that a critical $W_N$ string is related in the
sense described above, to a series
of $(N,N+1)$ $W_k$ minimal models with $k=2,3,...N-1$ \cite{PopeWN,own3}.
This is a special case of the situation
for non-critical $W_N$ strings (see e.g. \cite{BLNW}),
which may be related to more general minimal models \cite{BBRT}.

Most of the explicit results on $W$-strings are restricted to the $W_3$
string for which the BRST operator is known for some time \cite{TM}.
In this paper we go one step beyond $W_3$, i.e. we investigate the
physical spectrum of the critical $W_4$ string.
Recently, the $W_4$ BRST operator has been calculated \cite{W4BRST,own3}.
We will use here the operator obtained in \cite{own3}, since it has
a relatively simple structure.
In the basis that we use, the BRST operator can be decomposed into
separately nilpotent parts thus enabling the computation of the physical
operators in succesive steps. More explicitly, the $W_4$ BRST charge $Q$
contains a separately nilpotent BRST charge $Q_1$, associated to
spin-three and spin-four generators (thus excluding the Virasoro part of
the BRST charge), which in turn contains another BRST charge $Q_2$
associated to the spin-four generator only. Schematically, we have
\bea
\label{embedding}
Q_2 \subset Q_1 \subset Q\,.
\eea

We start to analyse the $Q_2$ cohomology which turns out to contain
all operators of the $(p,p')=(4,5)$ $W_3$ minimal model with central
charge $c_4^4={4\over5}$, realised in
terms of one scalar field and the spin-four ghost pair. In fact, there
seems to be an infinite number of representatives of
each minimal model primary.
They occur at different ghost numbers and are
connected by certain screening operations.
Next, the $Q_1$ cohomology turns out to contain all operators of the
$(4,5)$ Virasoro minimal model with central charge $c_4^3={7\over10}$.
Here again there seems to be an infinite number of copies of
each minimal model
primary at different ghost numbers.
In the final $Q$ cohomology all constraints are imposed, and all
physical operators are dimension zero primaries.
We try to sketch an overall picture of the different cohomologies
by applying the methods of \cite{PopeC}, i.e. by looking for special
physical operators that have physical inverses and can
normal order with any physical operator.

The strategy of computing the cohomology associated to a nilpotent
part of the BRST charge was followed in \cite{Hull} for
the $W_3$ string. In that case a nilpotent BRST charge is associated
to the spin-three constraint \cite{redef,own2}, and its cohomology
corresponds to the Ising model.

Some results on $W_4$ spectra were already obtained in \cite{DDR,PopeWN},
using not the BRST operator, but a certain
correspondence principle \cite{DDR}.
In particular, the connections with minimal models were already
noticed there. The relation with the $c={4\over5}$ $W_3$ minimal model
has been worked out further in \cite{PopeH2}.
In this paper we reproduce and extend these results using the full power
of the BRST operator.
We will also compare our results
with the cohomology classification of \cite{BMP2}.

\section{The $W_4$ string}
In order to study the physical spectrum of the $W_4$ string we need the
BRST operator for the $W_4$ algebra, which was given in \cite{W4BRST}.
A more convenient form of the BRST operator was found in \cite{own3}.
Let us summarize how it was obtained.

A three-scalar realisation of the $W_4$ algebra can be obtained from the
quantum Miura transformation for $su(4)$, see e.g. \cite{FL,DDR,PopeWN}.
In particular, the energy-momentum tensor is
\bea
\label{Memt}
T_M=-\h\partial\vec\phi\cdot\partial\vec\phi -\sqrt{2}\alpha_0
\vec\rho\cdot\partial^2\vec\phi\,,
\eea
where $\vec\phi =(\phi_1,\phi_2,\phi_3)$,
$\vec\rho$ is the Weyl vector
of $su(4)$ and $\alpha_0$ is a parameter
that will be specified later.
We use the representation
$\vec\rho={1\over2}(\sqrt{2},\sqrt{6},\sqrt{12})$,
and denote by $q_i =\sqrt{2}\alpha_0\rho_i$, $i=1,2,3$,
the background charges
of the scalar fields.
The central charge is
\bea
c_M=3+24(\alpha_0)^2\rho^2=3+120(\alpha_0)^2\,.
\eea
In a certain classical limit of this realisation (or the analogous
realisation of any $W_N$ algebra),
it is possible to redefine the generators
such that the algebra is brought to a special form with a nested
subalgebra structure \cite{own3}.
This means for a general $W_N$ algebra that the $k$ highest spin
currents form a subalgebra for any $k=1,2,...,N-1$.
This subalgebra structure arises due to the fact
that after the redefinition the highest spin current only depends on a
single scalar field, the next highest spin current only on this
plus an extra scalar field, etc.
The \emt\ is not affected by the redefinition.
The BRST charge associated to the resulting
classical algebra inherits the same nested structure. Quantisation by
parametrising
all possible quantum corrections and demanding nilpotency leads,
in the case of $W_4$, to a
BRST operator which still has this nested structure, and therefore, as we
will see, is very convenient for studying the spectrum.
Since we will make extensive use of this BRST operator we write it here
explicitly:
\bea
j_2 &=& c_4\{(\partial\phi_3)^4 + 4q_3\partial^2\phi_3
(\partial\phi_3)^2 + {\ts {41\over 5}}(\partial^2\phi_3)^2
+ {\ts {124\over 15}}\partial^3\phi_3\partial\phi_3 \nonumber\\
& &+{\ts {46\over 135}}q_3\partial^4\phi_3\}
\label{j44}
- 8(\partial\phi_3)^2c_4\partial c_4 b_4
+ {\ts {16\over 9}}q_3\partial^2\phi_3
c_4\partial c_4b_4 \\
& &+{\ts {32\over 9}}q_3\partial\phi_3c_4\partial^2c_4b_4
+{\ts {4\over 5}}c_4\partial^3c_4b_4 - {\ts {16\over 3}}c_4\partial c_4
\partial^2b_4\,,\nonumber\\
j_1 &=&
c_3\{(\partial\phi_2)^3 +
{\ts {3\over 4}}\partial\phi_2(\partial\phi_3)^2 +
{\ts {5\sqrt{2}\over 8}}(\partial\phi_3)^3
+ 3q_2\partial\phi_2\partial^2\phi_2\nonumber\\ & &+ {\ts {3\over2}}q_3
\partial\phi_2\partial^2\phi_3 + {\ts {9\over2}}q_2
\partial\phi_3\partial^2\phi_3
+ {\ts {93\over 40}}\partial^3\phi_2
+ {\ts {69\sqrt{2}\over 10}}\partial^3\phi_3\}\nonumber\\
& &- {\ts {9\over 2}}\partial\phi_2c_3\partial c_3b_3
+ {\ts {3\over2}}q_2 c_3\partial^2c_3b_3
- {\ts {243\over 64}}c_3\partial c_3b_4\label{j43}\\
& &- {\ts {9\over 2}}\partial\phi_2c_3c_4\partial b_4 -
6\partial\phi_2c_3\partial c_4b_4
+ {\ts {9\over2}}q_2 c_4\partial^2c_3b_4\nonumber\\
& &+ {\ts {3\over2}}q_2 c_3\partial^2c_4b_4 - {\ts {9\sqrt{2}\over 2}}
\partial\phi_3
c_4\partial c_3b_4 - 3\sqrt{2}\partial\phi_3c_3\partial c_4b_4
 + j_2 \,,\nonumber \\
j &=&
 c_2 (T_M + T_{c_3,b_3} + T_{c_4,b_4} + \h T_{c_2,b_2}) + j_1\,.
\label{j42}
\eea
Here $(c_k,b_k)$ is the conjugate ghost pair of the spin-$k$ symmetry with
conformal dimension $(1-k,k)$ with respect to the corresponding
\emt $\ T_{c_k,b_k}=-kb_k\partial c_k +(1-k)\partial b_k c_k$.
The total $W_4$ BRST charge is
$Q=\oint {dz\over2\pi i}j(z)$, and as the way of
representing it in eqs. (\ref{j44})-(\ref{j42}) suggests,
it involves two other nilpotent BRST charges:
$Q_2=\oint {dz\over 2\pi i}j_2(z)$ is a BRST operator corresponding
to a spin-4 symmetry, and $Q_1=\oint {dz\over 2\pi i}j_1(z)$ is a
BRST operator corresponding to a symmetry generated by spin-3 and spin-4
currents. We have
\bea
\label{alg}
(Q_2)^2=(Q_1)^2=(Q)^2=0\ \,;\ \ \ (Q_{Vir})^2=\{Q_{Vir},Q_1\}=0\,,
\eea
where we defined $Q_{Vir}=Q-Q_1$.
Note that $Q_{Vir}$ is just the usual Virasoro BRST operator for
the matter plus spin-(3,4) ghost systems.
The BRST operator in this special basis
is the generalisation to $W_4$ of the
$W_3$ BRST operator of \cite{redef}, which was obtained
from the original one \cite{TM} by
a canonical redefinition of the fields.
The spin-four part $Q_2$ was in fact already given in \cite{PopeH1}.
We should also mention that momenta
and ghost numbers of physical operators
are not affected by the redefinition leading to (\ref{j44}-\ref{j42}).
However, the explicit epressions of physical operators are expected to be
much simpler in the new basis, as was the case for the $W_3$ string
in \cite{redef}.

The BRST operator (\ref{j42}) is nilpotent
provided the total central charge
of matter plus ghosts vanishes. This requires $T_M$ to have
central charge $c_M=246$ implying $(\alpha_0)^2={81\over40}$.
Then we obtain what we call a critical $W_4$ string.
For a non-critical string (matter coupled to 2D gravity), one would expect
another sector with $W_4$ symmetry. Unfortunately,
however, the redefinition described above can only be applied to one of
both sectors \cite{own3} which means that the
description of the non-critical $W_4$ string
would be much more involved. On the other hand, in the non-critical case
one has the freedom to choose the partition of the central charge
over matter and gravity sectors, so that more general spectra of
physical states may be obtained.

Screening operators play an important role in the physical state
analysis. The standard $W_4$ screening currents \cite{FL} are
\bea
\label{CG}
S_i^{\pm}=e^{i\alpha_{\pm}\vec e_i\cdot\vec\phi}\,,
\eea
where $\vec e_i$ are the $su(4)$ simple roots, and
$\alpha_{\pm}$ are determined from the requirement that
the currents are spin-one primaries,
\bea
\label{alfa}
\alpha_{\pm}={\ts {i\over\sqrt{2}}}(\alpha_0\pm\sqrt{\alpha_0^2-2})
\eea
The screening charges $\oint {dz\over 2\pi i}S_i^{\pm}$ commute
with the $W_4$ generators in the
Miura realisation, and they
will appear in modified form in the
discussion of the total cohomology.
Besides, there will be more screening operators that simplify
the classification of physical states.
In general, for a physical operator ${\cal O}$
of zero conformal weight, one
can find an associated screening current $S_{\cal O}$ via the
descent equation
\bea
\label{descent}{}
[Q,S_{\cal O}(z)]=\partial {\cal O}(z)\,,
\eea
where $Q$ is the BRST charge under consideration.
The corresponding screening charge
$\oint {dz\over2\pi i} S_{\cal O}(z)$ will
then commute with $Q$.

Also important in the subsequent discussion are the picture changing
operators, one for each scalar field. They are defined by
\bea
P_i(z)=[Q,\phi_i(z)]\,,
\eea
and are not considered BRST trivial since the $\phi_i$ don't
belong to the set of conformal fields from which the (physical)
operators are constructed. Applying a picture changing operator
to a physical state, i.e. taking the normal ordered product,
either gives zero or another physical state.

An interesting selection rule for the momenta of scalar fields
with background charges was suggested in \cite{Rama}.
According to this rule, the momenta of physical states must be integral
multiples of the momenta of the screening currents (\ref{CG}):
\bea
\label{SR}
\vec p=\sum_{i=1}^{3} (n_i^+\alpha_+ + n_i^-\alpha_-)\vec e_i\,.
\eea
All physical operators described below indeed have momenta on the lattice
defined by (\ref{SR}).
It is then convenient to rewrite the momenta as
\bea
\label{p0}
p_i={i q_i\over27}\tilde p_i\,,
\eea
because (\ref{SR}) now implies that
\bea
\label{lattice}
\tilde p_1\in 3{\bf Z}\,;\ \ \tilde p_2\in {\bf Z}\,;\ \ \tilde p_3
\in 2{\bf Z}\,.
\eea
Note that the momenta (\ref{p0}) are imaginary.
In the following we will usually refer to $\tilde p_i$ as the momentum.

We will go on to determine the cohomology of $Q$ in steps, starting
with $Q_2$, which imposes the spin-four constraint.

\section{The $Q_2$ cohomology}
As the BRST current $j_2$ only depends
on the single scalar field $\phi_3$
and the spin-4 ghost pair $(c_4,b_4)$, we only need to consider operators
built from these fields.
These fields together have central charge $4\over5$, the central charge
of the $(p,p')=(4,5)$ $W_3$ unitary minimal model.

The $Q_2$ physical states at some low energy levels
were essentially already obtained
in \cite{PopeH1,PopeH2}, in a discussion of the spin-2 plus spin-4 string.
The extra Virasoro constraint included there, only seems
to dress the primary $Q_2$ physical operators
to operators of total spin zero.
Moreover, in the recent paper \cite{PopeW24}, the complete cohomology
of the critical spin-2 plus spin-4 string is given. We compare
with those results at the end of this section.

The ghost vacuum is given by acting on the
$sl_2$-invariant vacuum with $\partial^2 c_4\partial c_4c_4$.
In the following, we will always write down operators that are
supposed to act on the $sl_2$-invariant vacuum.
First consider operators of the form
\bea
\label{Q2l0}
V_0^0=\partial^2 c_4\partial c_4c_4\VX3\,.
\eea
The lower index $0$ stands for level $0$, the level being defined as the
conformal dimension of the operator in front of the exponential
minus the conformal dimension of the ghost vacuum (which is $-6$
since $c_4$ has conformal dimension $-3$).
The upper index in (\ref{Q2l0}) refers to the
ghost number $G$, which we define
to be $-3$ for the $sl_2$-invariant vacuum.
Level $1$ states at lowest ghost number ($G=-1$) take the form
\bea
\label{Q2l1}
V_1^{-1}=\partial c_4c_4\VX3\,.
\eea
The physical states of lowest ghost number at a particular level are easy
to find. Since they can't be $Q_2$-exact, one only
has to impose the vanishing of their $Q_2$-variation.
We will restrict ourselves to operators on levels $0$ and $1$, since
this will turn out to be enough to understand the global structure
of the $Q_2$ cohomology.
Imposing the physical state conditions on (\ref{Q2l0}) and (\ref{Q2l1}),
we obtain the results listed in table 1 a. and b., respectively.

\vspace{.1cm}

\renewcommand{\arraystretch}{1.1}
\begin{center}a.\
\begin{tabular}{|l|l|l|l|}
\hline
$V_0^0$  &  $\tilde p_3$  &  $h$  &  $w$ \\\hline
         &  $24$  &   $0$   &   $0$  \\\cline{2-4}
         &  $30$  &   $0$   &   $0$  \\\cline{2-4}
         &  $26$  &  ${1/15}$ & $1$ \\\cline{2-4}
         &  $28$  &  ${1/15}$ & $-1$ \\\hline
\end{tabular}
\hspace{1.3cm}b.\
\begin{tabular}{|l|l|l|l|}
\hline
$V_1^{-1}$ & $\tilde p_3$ & $h$ & $w$ \\\hline
           & $16$ & ${1/15}$ & $-1$ \\\cline{2-4}
           & $18$ & ${2/5}$  & $0$  \\\cline{2-4}
           & $20$ & ${2/3}$  & $-26$ \\\hline
\end{tabular}
\renewcommand{\arraystretch}{1}

\vspace{.5cm}

\parbox{10cm}{
\noindent {\bf Table 1.}\ \ \
{\it Level 0 and 1 physical states in
the $Q_2$ cohomology.
Momenta are denoted by $\tilde p_3$, see (\ref{p0}).
The last two columns give the weights $h$ and $w$ with respect to the
spin-two and three generators of the $c={4\over5}$ $W_3$ algebra.}}
\end{center}

\vspace{.4cm}

Note that the physical values of $p_3$ agree with eqs.
(\ref{SR})-(\ref{lattice}) (where we only have the third component of
(\ref{SR})).
The last two columns give the weights of the physical states
with respect to the generators of a $W_3$ algebra.
It turns out that the physical states in the $Q_2$ cohomology
can be organised in representations of the $c={4\over5}\,$ $W_3$ algebra
whose generators $(T,W)$ are physical operators
at levels $8$ and $9$ with
zero momentum. The Virasoro generator $T$ is just the \emt\ of the fields
$(\phi_3,c_4,b_4)$, and the spin-three generator is given
by \cite{PopeH2}
\bea
W&=&\sqrt{{\ts {2\over13}}}\{{\ts {5\over3}}(\dphi_3)^3 +
5q_3\partial^2\phi_3\dphi_3
+{\ts {25\over4}}\partial^3\phi_3 + 20\dphi_3 b_4\partial c_4\nonumber\\
& &+12\dphi_3\partial b_4 c_4
+12\d2phi_3 b_4 c_4 +5q_3\partial b_4
\partial c_4 +3q_3\partial^2 b_4 c_4\}\,.
\label{W3}
\eea
In fact they generate the $c={4\over5}\ W_3$ algebra
with standard normalization up to an extra primary spin-four
operator, which turns out to be a multiple of the spin-four
BRST-exact generator $V=\{Q_2,b_4\}$.
It was noticed by the authors of \cite{PopeH2} that after bosonising the
spin-four ghost pair, this realisation of the $W_3$ algebra coincides
with a special two-scalar realisation found in \cite{own1}.
In \cite{own1} a systematic investigation of free-scalar realisations of
the $W_3$ algebra was performed, in which
the OPE of the spin-three generators
was allowed to contain a norm zero spin-four operator. Among all the
two-scalar realisations this special $c={4\over5}$ realisation, obtained
after bosonising the spin-four ghosts in (\ref{W3}), is unique in the sense
that it has one real background charge and one imaginary background charge,
the latter belonging to the ``ghost scalar''.

The physical states in table 1 are all primary with respect to
the $W_3$ algebra
and their $L_0$ and $W_0$ eigenvalues are denoted by
$h$ and $w$, respectively. For convenience the $W$ weights $w$ have
been rescaled as in \cite{PopeH2}.
The Virasoro weights $h$ are given in terms of $p_3$ as
\bea
\label{weight}
h=\h (p_3)^2 - i q_3 p_3 +l-6\,,
\eea
where $l$ is the level. The spin-three weight is a cubic polynomial
in $p_3$, and depends on the detailed structure of the operator.

Let us now compare the $Q_2$ spectrum with the spectrum of
primaries in a $c={4\over5}\ W_3$ minimal model.
The spectrum of conformal weights in a generic $(p,p')$ $W_3$ minimal
model is given by (see e.g. \cite{BS})
\bea
\label{W3Kac2}
h(r_1,r_2;s_1,s_2)&=&-{(p-p')^2\over pp'}\\
& &\hspace{-2.5cm}+{1\over 3pp'}\{\sum_{i\leq j=1}^2
(p'(r_i+1)-p(s_i+1))(p'(r_j+1)-p(s_j+1))\}\,,\nonumber
\eea
where the non-negative integers $r_i,s_i$ run over the range
\bea
0\leq r_1+r_2\leq p-3\,;\ \ 0\leq s_1+s_2\leq p'-3\,.
\eea
Note that the level 0 states in table 1 correspond to the ``diagonal''
entries of the $(p,p')=(4,5)$ Kac table (\ref{W3Kac2}), since
$h(0,0;0,0)=0$ and $h(0,1;0,1)=h(1,0;1,0)={1\over15}$.
The weights ${2\over5}$ and ${2\over3}$ of the level 1 states
are also in the set (\ref{W3Kac2}),
moreover, at levels 0 and 1 together, all
conformal weights of the
$(4,5)\ W_3$ minimal model occur.
It is also interesting to note that
the maximum possible conformal dimension
of an operator at a particular level, $h_{max}={1\over2}q_3^2 +l-6$
(see (\ref{weight})),
forbids the appearance of e.g. $h={2\over5}$ or $h={2\over3}$
operators on level 0.

\noindent The spin-three weights $w$ corresponding to the
Virasoro weights $h(r_1,r_2;s_1,s_2)$ in a $(p,p')\ W_3$ minimal model
are given by \cite{BS}
\bea
\label{W3Kac3}
w(r_1,r_2;s_1,s_2)&=&C(p,p') (p'(r_1-r_2)-p(s_1-s_2))\\
& &\hspace{-4cm}\times
(p'(2r_1+r_2+3)-p(2s_1+s_2+3))(p'(r_1+2r_2+3)-p(s_1+2s_2+3))\,,
\nonumber
\eea
where $C(p,p')$ depends on the normalization of the spin-three current.
The $w$-values in table 1 are indeed in agreement with the minimal model
values (\ref{W3Kac3}).
Under the ${\bf Z}_2$ transformation $(r_1,r_2;s_1,s_2)\rightarrow
(r_2,r_1;s_2,s_1)$ the $h$ values are invariant, while the $w$ values
change sign. We observe that the level 0 states occur in these
${\bf Z}_2$ pairs.
Looking at table 1, it is clear that a physical state with $(h,w)=
({2\over3},+26)$ is missing on levels 0 and 1. However, we only discussed
states of lowest ghost number. In particular, any state of ghost
number $G$ has a conjugate state at ghost number $1-G$ at the same level,
and it turns out that the $({2\over3},+26)$ state
occurs at level 1, $G=2$.
It is in fact the conjugate of the $({2\over3},-26)$ state in table 1.b.
More generally, conjugation seems to be associated to the ${\bf Z}_2$
symmetry mentioned above, since $h$ is invariant and $w$ changes sign
under conjugation.
This completes the identification of all minimal model primaries
in the $Q_2$ cohomology, at levels 0 and 1.

We now want to study the spectrum of $Q_2$ physical states more
thoroughly. For that purpose we introduce the following
screening operators,
\bea
\label{S3}
S&=&b_4 \VX3\,,\ \ \ \ \ \ {\rm with}
\ \tilde p_3=-6\,,\\
\label{R3}
R&=&\partial c_4c_4 \VX3\,,
\ \ \ \ \ \ \ \ \ \tilde p_3=30\,,\\
\label{R3bar}
\bar R&=&\partial c_4c_4 \VX3\,,
\ \ \ \ \ \ \ \ \ \tilde p_3=24\,.
\eea
They are spin-one primaries whose charges commute with $Q_2$.
It is not difficult to see that $R$ and $\bar R$ are
the screening currents associated to the
level 0, $h=0$ physical operators of table 1.a via the descent
equation (\ref{descent}).
With these screening charges it is possible to obtain new physical
states by acting on e.g. the level 0 and 1 states described above.
The OPE's of $T$ and $W$ with the screening currents
are total derivatives
(in the case of $R$ and $\bar R$ this is true up to $Q_2$-exact terms),
which means that $W_3$ primaries are mapped to $W_3$ primaries
of the same $(h,w)$ by the action of the screening charges.

We now apply the methods of
\cite{FWcoh} that were used in
a similar discussion for the $W_3$ string.
For the action of $n$ screening charges
on a physical state of momentum $p$ to be well-defined, the
following expression must be an integer \cite{FWcoh},
\bea
\label{Pn}
P_n\equiv n-1 +\sum_{i<j=1}^n p_{s_i}p_{s_j}+p\sum_{i=1}^n
p_{s_i}\,,
\eea
with screening momenta $p_{s_i}$.
Using this, one can show that e.g. the action of $S$ on
$V_0^0[\tilde p_3=30]$ is well-defined.
However, this action\footnote{By the action of a screening operator $S$
on a physical operator $V$ we mean the commutator $\oint {dz\over 2\pi i}
S(z)V(w)$,
whereas the action of a picture changing
operator $P$ is the normal ordered
product $\oint {dz\over 2\pi i}{P(z)V(w)\over z-w}$,
with integration contour around $w$.}
is trivial in the
sense that it gives zero, and to obtain a new physical state, we have
to make use of the picture changing operator $P_3(z)=[Q_2,\phi_3(z)]$.
Taking the normal ordered product of $P_3$ with $V_0^0[\tilde p_3=30]$
and then
acting with $S$ gives the physical state $V_0^0[\tilde p_3=24]$.
Generalising this, we can write down infinite series of operators
by analogy with the $W_3$ case \cite{FWcoh,Hull}.
Defining $V(0,0)=V_0^0[\tilde p_3=30]$, the series with $(h,w)=(0,0)$
may be written as
\bea
\label{hw00}
\bar V(0,n)=S P_3 V(0,n)\,,\ \ V(0,n)=(S)^4 P_3 \bar V(0,n-1)\,.
\eea
The other series are
\bea
&&V_-({\ts {1\over15}},0)\equiv V_0^0[\tilde p_3=28]\,,\\
\bar V_-({\ts {1\over15}},n)&=&(S)^2 P_3 V_-({\ts {1\over15}},n)\,,
\ \ V_-({\ts {1\over15}},n)=(S)^3 P_3 \bar V_-({\ts {1\over15}},n-1)\,;
\nonumber\\
&&V_+({\ts {1\over15}},0)\equiv V_0^0[\tilde p_3=26]\,,\\
\bar V_+({\ts {1\over15}},n)&=&(S)^3 P_3 V_+({\ts{1\over15}},n)\,,
\ \ V_+({\ts {1\over15}},n)=(S)^2 P_3 \bar V_+({\ts{1\over15}},n-1)\,;
\nonumber\\
&&V({\ts{2\over5}},0)\equiv V_1^{-1}[\tilde p_3=18]\,,\\
\bar V({\ts{2\over5}},n)&=&(S)^2 P_3 V({\ts{2\over5}},n)\,,
\ \ V({\ts{2\over5}},n)=(S)^3 P_3 \bar V({\ts{2\over5}},n-1)\,;
\nonumber\\
&&V_-({\ts{2\over3}},0)\equiv V_1^{-1}[\tilde p_3=20]\,,\\
\bar V_-({\ts{2\over3}},n)&=&S P_3 V_-({\ts{2\over3}},n)\,,
\ \ V_-({\ts{2\over3}},n)=(S)^4 P_3 \bar V_-({\ts{2\over3}},n-1)\,;
\nonumber\\
\label{hw23p}
&&V_+({\ts{2\over3}},0)\equiv V_1^1[\tilde p_3=34]\,,\\
\bar V_+({\ts{2\over3}},n)&=&(S)^4 P_3 V_+({\ts{2\over3}},n)\,,
\ \ V_+({\ts{2\over3}},n)=S P_3 \bar V_+({\ts{2\over3}},n-1)\,.
\nonumber
\eea
The notation, not to be confused with the previous notation $V_l^G$
with level and ghost number indices,
is $V_{\pm}(h,n)$, where $h$ is the spin and
$\pm$ indicates the sign of the spin-three weight $w$ (see table 1).
Although the actions of the screening operator $S$ in general
do not seem to have inverses, one can act on any operator in
(\ref{hw00})-(\ref{hw23p}) with $R$, thus extending
the series to negative $n$
\bea
V_{\pm}(h,n-1)=R P_3 V_{\pm}(h,n)\,,\nonumber\\
\bar V_{\pm}(h,n-1)=R P_3 \bar V_{\pm}(h,n)\,.
\eea
To summarize, we list all these operators
with their momentum, ghost number
and level at which they occur in table 2.

\vspace{.1cm}

\renewcommand{\arraystretch}{1.3}
\begin{center}
\begin{tabular}{|l|l|l|l|}
\hline
operator & $\tilde p_3$ & $G$   & level          \\\hline
$V(0,n)$ & $30-30n$    & $-3n$ & $\h(3n(5n-1))$ \\\hline
$\bar V(0,n)$ & $24-30n$ & $-3n$ & $\h(3n(5n+1))$ \\\hline
$V_-({\ts {1\over15}},n)$ & $28-30n$ & $-3n$ & $\h(n(15n-1))$ \\\hline
$\bar V_-({\ts {1\over15}},n)$ & $16-30n$ & $-1-3n$ & $\h(15n^2+11n+2)$
\\\hline
$V_+({\ts{1\over15}},n)$ & $26-30n$ & $-3n$ & $\h(n(15n+1))$ \\\hline
$\bar V_+({\ts{1\over15}},n)$ & $8-30n$ & $-2-3n$ & $\h(15n^2+19n+6)$
\\\hline
$V({\ts{2\over5}},n)$ & $18-30n$ & $-1-3n$ & $\h(15n^2+9n+2)$ \\\hline
$\bar V({\ts{2\over5}},n)$ & $6-30n$ & $-2-3n$ & $\h(15n^2+21n+8)$ \\\hline
$V_-({\ts{2\over3}},n)$ & $20-30n$ & $-1-3n$ & $\h(15n^2+7n+2)$ \\\hline
$\bar V_-({\ts{2\over3}},n)$ & $14-30n$ & $-1-3n$ & $\h(15n^2+13n+4)$
\\\hline
$V_+({\ts{2\over3}},n)$ & $34-30n$ & $1-3n$ & $\h(15n^2-7n+2)$ \\\hline
$\bar V_+({\ts{2\over3}},n)$ & $10-30n$ & $-2-3n$ & $\h(15n^2+17n+6)$
\\\hline
\end{tabular}

\renewcommand{\arraystretch}{1}

\vspace{.5cm}

\parbox{10cm}{
\noindent {\bf Table 2.}\ \ \ \
{\it Operators in the $Q_2$ cohomology.}}
\end{center}

\vspace{.5cm}

It would be interesting to see whether these operators are all
BRST non-trivial (they are certainly BRST closed).
In fact, the operators in table 2 are
supposed to be so-called prime operators
\cite{Popediscr}.
{}From them, new physical operators can be obtained by normal ordering
with the picture changing operator $P_3$, so all operators come
in doublets. The states in a doublet can be viewed as states
built on different degenerate vacua,
the degeneracy being caused by the ghost zero-modes,
just as in the ordinary string.
We observe that the operator $V(0,1)$ is the identity, and $\bar V(0,1)$
is another $h=0$ operator at the same ghost number $G=-3$ relative to
the tachyonic operators (\ref{Q2l0}).

{}From (\ref{hw00})-(\ref{hw23p}) one observes that the action of five
$S$ screening charges (together with two picture changes) is special.
It lowers $\tilde p_3$ by $30$ and $G$ by $3$. Indeed, a
screening operator exists which does the same in one go
(together with one picture change), namely
\bea
\label{Sx}
S_x=\partial^3 b_4\partial^2 b_4\partial b_4b_4 \VX3
\,,\ \ {\rm with}\ \tilde p_3=-30\,.
\eea
This screening operator is also used in the recent paper \cite{PopeW24}.
On all operators considered, $S_x P_3$ is the inverse action of $R P_3$.
The physical operator $x$ associated to $S_x$ can be constructed
using (\ref{descent}). It turns out to be equal (up to an irrelevant
constant factor) to
\bea
\label{xS}
x(z)=\oint {dw\over 2\pi i} S_x(w)P_3(z)\,,
\eea
which is precisely the operator $V(0,2)$ from (\ref{hw00}).
We do not write down $x$ explicitly, since it is a complicated
expression with 50 terms.
Conversely, $S_x$ is reobtained from $x$ as
\bea
\label{Q2descent}
S_x(w)=(\hat b_4)_{-1}x(w)\equiv\oint
{dz\over 2\pi i} (z-w)^2 b_4(z)x(w)\,.
\eea
Just as is the case in \cite{PopeC} for $W_3$,
$x$ has a physical inverse operator,
$x^{-1}$, such that the normal ordered product of $x$ with
$x^{-1}$ is a non-vanishing multiple of the identity. This inverse
is precisely the physical level 0 operator $V_0^0[\tilde p_3=30]$.
We may write it as
\bea
\label{xiR}
x^{-1}=\oint {dw\over 2\pi i}R(w)P_3(z)\,.
\eea
So we have identified three members of the family (\ref{hw00}):
$V(0,0)=x^{-1}$, $V(0,1)={\bf 1}$ and $V(0,2)=x$.

In \cite{PopeC} the invertibility of $x$ enabled
the computation of the complete cohomology of the critical $W_3$ string
by the observation that normal ordered products of arbitrary powers of
$x$ or $x^{-1}$ with a physical operator give new
non-trivial physical operators.
The situation here is somewhat different though. The reason is that
the operators in the $Q_2$ cohomology do not all have total conformal
weight zero, due to the lack of a Virasoro constraint.
Due to this, we have not been
able to prove that the operators of table 2
generate the complete $Q_2$ cohomology.

If we assume for the moment that all operators of table 2 are prime
physical operators, then
using (\ref{xS}) and (\ref{xiR}), it
can be shown that the combined action
of $S_x$ and $P_3$  on a prime physical operator is
equivalent to normal ordering with $x$, and similarly for $R P_3$ and
$x^{-1}$.
It is therefore quite likely that the operators of table 2 together with
their $W_3$ descendants
and their picture changed versions form the complete $Q_2$ cohomology.

Equation (\ref{Q2descent}) seems to give the general procedure
to obtain the screening current
associated to a $h=0$ physical operator in the $Q_2$ cohomology.
For $S_{{\cal O}}\equiv (\hat b_4)_{-1}{\cal O}$, with ${\cal O}$
an arbitrary $h=0$ physical operator, one has
\bea
{}
[Q_2,S_{{\cal O}}(w)]=\hat V_{-1}{\cal O}(w)=\oint {dz\over 2\pi i}
(z-w)^2 V(z) {\cal O}(w)\,,
\eea
where as before $V$ is the spin-four current $\{Q_2,b_4\}$,
and the RHS is indeed a (multiple of) $\partial{\cal O}$ in the
cases we examined.

During the course of this work we received the paper \cite{PopeW24}
in which the complete cohomology of the critical $W_{2,4}$ string
is given. Ignoring the Virasoro constraint, i.e. ignoring the
``Liouville dressings'', this cohomology seems to be
equivalent to the $Q_2$ cohomology of highest weight operators
obtained above, apart from some extra descendants which couple to
the Virasoro fields in the $W_{2,4}$ cohomology.

\section{The $Q_1$ cohomology}
We now take the next nilpotent BRST operator $Q_1$ of
equation (\ref{j43}) and study its cohomology.
It is the part of the total $W_4$ BRST current (\ref{j42})
which does not involve the Virasoro sector. It imposes only
a spin-three and a spin-four constraint.
The Fock space must now be extended to include also the scalar $\phi_2$
and the spin-three ghost pair.
Together, the fields $(\phi_2,\phi_3;c_3,b_3;c_4,b_4)$ have
central charge $c={7\over10}$ which is the central charge of the
$(p,p')=(4,5)$ unitary Virasoro minimal model.

Operators in the $Q_1$ cohomology can be computed from operators
in the $Q_2$ cohomology in a systematic way
using a spectral sequence argument (for a review, see e.g \cite{BMP}).
Taking the spin-three ghost number $G_3$ as an extra grading
on the complex of scalar plus ghost Fock spaces, one can decompose
$Q_1$ in three parts with $G_3=0,1$ and $2$:
\bea
\label{spseq}
Q_1=d_0+d_1+d_2\,,
\eea
where the $G_3=0$ part, $d_0$, is just $Q_2$. There is only one term
in $Q_1$ which has $G_3=2$, namely $d_2=-{243\over64}c_3\partial c_3
b_4$. This term prevents the complex from being a double complex.
The rest has $G_3=1$ and is denoted by $d_1$ in (\ref{spseq}).
The first term of the spectral sequence
$(E_r,\delta_r)_{r=0}^{\infty}$ associated to this gradation,
is the $Q_2$ cohomology
\bea
E_1&=&H(Q_2,{\cal F}(\phi_2,\phi_3;c_3,b_3;c_4,b_4))\nonumber\\
&=&
{\cal F}(\phi_2;c_3,b_3)\otimes  H(Q_2,{\cal F}(\phi_3;c_4,b_4))\,,
\eea
where the second equality follows from the fact that $Q_2$ does not
involve any of the fields $(\phi_2;c_3,b_3)$.
So we can start with a $Q_2$ physical operator and extend it
(if possible) to a $Q_1$ physical operator by computing the
next terms in the spectral sequence.
The successive terms that are added
to the original $Q_2$ physical operator
have increasing $G_3$ value (but of course the same total ghost number).
On the lower lying levels the spectral sequence will collapse
after a few terms due
to the small range of ghost numbers availabe there, but at higher
levels the procedure becomes increasingly laborious.

Having said this, we found it just as convenient
to compute operators in the $Q_1$ cohomology by imposing
the complete $Q_1$ physical condition at
once. However, it is very useful
to observe from the arguments mentioned above, that the $Q_1$
physical operators are extensions of $Q_2$ physical operators, so
that only the $\phi_2$ momentum
and the spin-three ghost structure remain to be determined
from the $Q_1$ physical condition.

The level 0 operators are now of the form
\bea
\label{Q1l0}
W_0^0=\partial c_3c_3\partial^2 c_4\partial c_4c_4 e^{ip_2\phi_2
+ip_3\phi_3}\,.
\eea
The notation is the same as in (\ref{Q2l0}) except that we use $W$
for operators in the $Q_1$ cohomology.
Level 1 operators at lowest ghost number $G=-1$ can now be
linear combinations of two terms with different ghost structure:
\bea
\label{Q1l1}
W_1^{-1}=(x_1c_3\partial^2 c_4\partial c_4c_4+x_2\partial c_3c_3
\partial c_4c_4)e^{ip_2\phi_2+ip_3\phi_3}\,.
\eea
Tables 3 a. and b. list the momenta for which the level 0
and level 1 operators, respectively, are physical.

The cohomology classes corresponding to the given momenta
are all one-dimensional. The level 1 operators at $\tilde p_3$-momenta
$16,18$ and $20$ have $x_1=0$ in (\ref{Q1l1}), the
other level 1 operators
have certain non-zero ${x_1\over x_2}$ ratios.

\vspace{.1cm}

\renewcommand{\arraystretch}{1.1}
\begin{center}a.\
\begin{tabular}{|l|l|l|l|}
\hline
$W_0^0$ & $\tilde p_3$ & $\tilde p_2$ & $h$ \\\hline
&24 & 24 & 0 \\\cline{3-4}
& & 27 & 3/80 \\\cline{3-4}
& & 30 & 0 \\\cline{2-4}
&26 & 22 & 0 \\\cline{3-4}
& & 28 & 1/10 \\\cline{3-4}
& & 31 & 3/80 \\\cline{2-4}
&28 & 23 & 3/80 \\\cline{3-4}
& & 26 & 1/10 \\\cline{3-4}
& & 32 & 0 \\\cline{2-4}
&30 & 24 & 0 \\\cline{3-4}
& & 27 & 3/80 \\\cline{3-4}
& & 30 & 0 \\\hline
\end{tabular}
\hspace{2cm}b.\
\begin{tabular}{|l|l|l|l|}
\hline
$W_1^{-1}$ & $\tilde p_3$ & $\tilde p_2$ & $h$ \\\hline
&16 & 23 & 3/80 \\\cline{3-4}
& & 26 & 1/10 \\\cline{3-4}
& & 32 & 0 \\\cline{2-4}
&18 & 18 & 1/10 \\\cline{3-4}
& & 27 & 7/16 \\\cline{3-4}
& & 36 & 1/10 \\\cline{2-4}
&20 & 19 & 7/16 \\\cline{3-4}
& & 22 & 3/5 \\\cline{3-4}
& & 40 & 0 \\\cline{2-4}
&24 & 12 & 1/10 \\\cline{3-4}
& & 15 & 7/16 \\\cline{2-4}
&26 & 16 & 3/5 \\\cline{2-4}
&28 & 11 & 3/80 \\\cline{2-4}
&30 & 12 & 1/10 \\\cline{3-4}
& & 15 & 7/16 \\\hline
\end{tabular}

\renewcommand{\arraystretch}{1}

\vspace{.5cm}

\parbox{10cm}{
\noindent {\bf Table 3.}\ \ \
{\it Level 0 and 1 operators in the $Q_1$ cohomology.}}
\end{center}

\vspace{.4cm}

The last column in table 3 gives the total conformal weight
of the physical operators with respect to the $c={7\over10}$
\emt, which is itself a physical operator at level 11.
Thus the physical states are organised into $c={7\over10}$
Virasoro representations. Unitarity would then imply that
all primary physical operators should have conformal dimensions of
the corresponding Kac table. This seems to be the case.
In particular, the level 0 physical operators correspond to
the diagonal entries of the Kac table.
The multiplicities of operators of fixed weight agree
with those predicted by the Weyl group approach in \cite{PopeWN}.
The presence of non-diagonal
operators at level 0 is impossible because of
the maximum conformal weight
\bea
\label{maxcd}
h_{max}(l)=\h(q_2^2+q_3^2)+l-9={\ts{9\over80}}+l\,.
\eea
At level 1, primary operators
corresponding to the first off-diagonal in the Kac table,
with $h={7\over16}$ and $h={3\over5}$, are
allowed by (\ref{maxcd}), and they are indeed in the $Q_1$ cohomology
as can be seen from table 3.b.
Also observe that there is no physical operator corresponding to
the outermost entry in the Kac table, $h={3\over2}$, at levels 0
and 1. From (\ref{maxcd}) it is clear that such an operator
can exist only at levels $l\geq 2$.
So it is natural to look for this missing operator on level 2.
At this level, the ghost number can take values $-2\leq G\leq4$.
To see if there is a $h={3\over2}$ physical state, it is enough
to consider only $G\leq 0$, since
for $G\geq 1$ the spectrum consists of
conjugates of $G\leq 1$ states with the same conformal weight.
The lowest ghost number operator at level 2 has the form
$W_2^{-2}=c_3\partial c_4c_4 e^{ip_2\phi_2+ip_3\phi_3}$, and
is physical for two values of the momenta $(p_2,p_3)$ giving rise
to two $h={3\over80}$ operators. At $G=-1$, there is no $h={3\over2}$
cohomology either. But at $G=0$
there is a one-dimensional $h={3\over2}$
cohomology class, with momentum $(\tilde p_2,\tilde p_3)=(34,20)$.
It may be represented by $\partial^3 c_3\partial c_3 c_3\partial c_4c_4
e^{ip_2\phi_2+ip_3\phi_3}$, which is primary up to $Q_1$-exact terms.
There is also a cohomology class with the
conjugate momentum\footnote{In general, the momentum conjugate to
$\vec p$ is $\vec p^C=2i\vec q -\vec p$,
where $\vec q$ is the background charge
vector. Using (\ref{p0}) this translates to $\tilde p_i^C=
54-\tilde p_i$.}
(and thus also $h={3\over2}$),
$(\tilde p_2,\tilde p_3)=(20,34)$.
One can understand the appearance of states at $G=0$ in pairs
with conjugate momenta as follows.
First note
that states in the $Q_1$ cohomology occur in quartets
with ghost numbers $(G,G+1,G+1,G+2)$, where
the state at lowest ghost number
is called the prime state \cite{Popediscr}, and the other states are
obtained by applying the picture changing operators to this prime state
(remember that we have two independent picture changing operators in
the $Q_1$ cohomology description). Besides, any state at ghost number
$G$ has a conjugate state at ghost number $2-G$ with
the conjugate momentum.
Therefore, at $G=0$, prime states occur in pairs with conjugate
momenta.

We have now identified all operators of the $c={7\over10}$ minimal model
in the $Q_1$ cohomology at levels 0,1 and 2. The next objective is
to show that all physical operators (at least the ones found so far)
of the same conformal weight are related to each other
through the action of screening operators and picture changes.
Let us introduce therefore a number of useful screening charges,
which are now required to commute with $Q_1$. First of all,
the operator $S$ of equation (\ref{S3}) is still a screening current
in the $Q_1$ cohomology, as is $S_x$ (eq. (\ref{Sx})).
The $Q_2$ screening operators
$R$ and $\bar R$ commute with $Q_1$ only after adding an extra term,
\bea
\label{R}
R&=&(\partial c_4c_4+{\ts{15\over88}}q_2c_3c_4)\VX3
\,,\ \ \tilde p_3=30\,,\\
\bar R&=&(\partial c_4c_4+{\ts{15\over56}}q_2c_3c_4)
\VX3\,,\ \ \tilde p_3=24\,.
\eea
New screening currents are given
by (the notation will become clear in the
next section)
\bea
\label{T3m}
T_3^-&=&(1-{\ts{256\over729}}q_2 b_3c_4)
\Vp\,,\ \ \
\tilde{\vec p}\equiv(\tilde p_2,\tilde p_3)=(-8,8)\,,\\
\label{T3p}
T_3^+&=&(1-{\ts{32\over729}}q_2 b_3c_4)
\Vp\,,\ \ \
\tilde{\vec p}=(-10,10)\,.
\eea
Screening currents with positive $\tilde p_2$-values are
\bea
R'&=&c_3 e^{i p_2\phi_2}\,,\ \ \tilde p_2=30\,,\\
\bar R'&=&c_3 e^{ip_2\phi_2}\,,\ \ \tilde p_2=24\,.
\eea
Of course, many more screening currents at higher or lower ghost numbers
exist, but we expect that they can be represented by composite actions
of the given ones, together with $P_2$ and/or $P_3$ picture changes.

The $h=0$ physical operators are obtained through the action
of the associated screening currents on a picture changed version of
the identity operator, so they can all be viewed as different
screened versions of the identity.
Also, operators of table 3 with the same conformal weight can be
connected more directly to each other by the action of certain
combinations of the screening charges given above.
More important, however, is to find
operators which can normal order with any physical operator and thereby
create new physical operators.
The operator $x$ that was found in the previous section, is easily
extended to the $Q_1$ cohomology, since the associated screening current
$S_x$ is still given by (\ref{Sx}).
It is again also given by
\bea
\label{SxQ1}
x(z)=\oint {dw\over 2\pi i} S_x(w)P_3(z)\,,
\eea
where now $P_3=[Q_1,\phi_3]$ contains some extra terms compared to
the $P_3$ operator in the $Q_2$ discussion.
There is again a physical inverse operator, $x^{-1}$. It is
the level 3 physical operator
\bea
x^{-1}&=&(\partial^2 c_4\partial c_4c_4+{\ts {45\over56}}\dphi_2 c_3
\partial c_4c_4-{\ts{45\sqrt{2}\over56}}\dphi_3 c_3\partial c_4c_4
-{\ts{5\over56}}q_2c_3\partial^2 c_4c_4\nonumber\\
& &+{\ts{5\over28}}q_2\partial c_3\partial c_4c_4
+{\ts{3645\over19712}}\partial c_3c_3c_4)\VX3\,,
\label{xinv1}
\eea
with momentum $\tilde p_3=30$. This is the operator
corresponding to the screening current (\ref{R}) via the descent
equation, but it is also equal to the commutator of $R$ with $P_3$.
Conversely, we get back $R$ as $(\hat b_3)_{-1} x^{-1}$.

There must also be similar operators $y$ and $y^{-1}$ with non-zero
$\phi_2$ momentum. We expect $y$ to be a level 37 physical operator
with momentum $(\tilde p_2,\tilde p_3)=(-40,-20)$. Such an operator
can normal order with any physical operator. This can be easily checked
using (\ref{Pn}), and by noting that $\tilde p_2 + \tilde p_3$ is a
multiple of 3 for all physical operators (this also follows
from (\ref{SR})).
We did not try to construct the operator $y$. However,
$y^{-1}$ should
just be the level 1 physical operator with momentum
$(\tilde p_2,\tilde p_3)=(40,20)$, see e.g table 3.
So there are strong indications that the operators $x,y,x^{-1},y^{-1}$,
generate the entire $Q_1$ cohomology in their action on a number of
physical operators at low-lying levels \cite{PopeC}.
Unfortunately, we cannot be more precise in these statements,
since that would, among other things, require
the explicit construction of
$y$ and the verification of $y^{-1}(z)y(w)=O(1)$.
The explicit construction of new physical operators by normal ordering
e.g. level 0 or 1 physical operators with powers of $x,y,x^{-1},y^{-1}$
would be quite complicated as well. However, since the ghost numbers
and momenta of the operators $x,y,x^{-1},y^{-1}$ are known, they may be
used to predict the ghost numbers, level numbers, and momenta
of $Q_1$ cohomology classes, just
like in table 2 for the $Q_2$ cohomology.

The overall picture of the $Q_1$ cohomology is then the following.
Physical operators come in minimal model modules of the $c={7\over10}$
Virasoro algebra, realised in terms
of the scalar fields $(\phi_2,\phi_3)$
and the ghost pairs $(c_3,b_3;c_4,b_4)$. There seems to be an infinite
number of representatives of each minimal model primary (but only a
finite number at fixed ghost number).  We expect that all primaries
belonging to the $Q_1$ cohomology can be written as normal ordered
products of powers of the operators $x,y$ and
their inverses acting on a set
of low level physical operators.

\section{The complete cohomology}
We now consider the cohomology of $Q=Q_1+Q_{Vir}$
on the full Fock space
generated by the three scalar fields and the three conjugate ghost pairs.
Their central charges add up to zero.
For $c=0$, the only unitary representation of the Virasoro algebra
has $h=0$. Indeed, the extra Virasoro constraint
imposed by $Q_{Vir}$ guarantees that all physical operators have $h=0$
with respect to the total \emt\ (\ref{Memt}).
Moreover, the $h=0$ representation is trivial, so the descendents are
null and BRST trivial in the cohomology. Thus the $Q$ cohomology contains
only $h=0$ primaries. Of course, this is just like the situation for
the ordinary bosonic string, but it is different from the $Q_2$ and $Q_1$
cohomologies where descendants of minimal model primaries are also
present.

Since $Q_1$ and $Q_{Vir}$ anti-commute, see (\ref{alg}), they define
a double complex. Note that $Q_1$ does not involve the fields $(\phi_1;
c_2,b_2)$, and
since $Q_1$ physical operators have already been computed,
we take a spectral sequence where the first term is the $Q_1$
cohomology. This spectral sequence provides a
systematic procedure to obtain
operators in the total cohomology
by adding to $Q_1$ physical operators terms
with higher spin-two ghost number
$G_2$ (but the same
total ghost number). Physical operators in the total cohomology are
then given by
\bea
\label{tic-tac-toe}
{\cal O}=\sum_{i=k}^{\infty}{\cal O}_{i}\,,
\eea
where the first term in the sum, ${\cal O}_k$ with $G_2=k$,
is an operator in the $Q_1$ cohomology, and the higher $G_2$ terms
are defined by $[Q_1,{\cal O}_{i+1}]=-[Q_{Vir},{\cal O}_i]$.
At small values of the level, the sum in (\ref{tic-tac-toe})
will only have a few terms (at level 0 and $G=0$ only one term).

The tachyonic operators (level 0, $G=0$) are of the form
\bea
X_0^0=c_2\partial c_3c_3\partial^2 c_4\partial c_4c_4
e^{ip_1\phi_1+ip_2\phi_2+ip_3\phi_3}\,.
\eea
They are physical for 24 values of the momenta, listed in table 4.a.
In fact, they are just
the 12 $Q_1$ operators of table 3.a dressed up with the $c_2$ ghost and
the $\phi_1$ part of the exponential, to operators of vanishing total
conformal dimension (thereby giving two possible $p_1$ values).
In \cite{DDR}, these level 0 operators were already
given. Their computation
was based on an assumption about the existence of the ``cosmological
constant operator''. In \cite{PopeWN}
it was shown that these momenta can be obtained by the action of elements
of the $su(4)$ Weyl group (of which there are $4!=24$) on
a specific solution.
We come back to this point at the end of this section.

At level 1 and lowest ghost number $G=-1$,
\bea
X_1^{-1}=(x_1\partial c_3c_3\partial^2 c_4\partial c_4c_4+
x_2c_2c_3\partial^2 c_4\partial c_4c_4+
x_3c_2\partial c_3c_3\partial c_4c_4)e^{i\vec p\cdot\vec\phi}\,,
\eea
which is physical for the momenta in table 4.b.

\vspace{.5cm}

\renewcommand{\arraystretch}{1.1}
\begin{center}a.\
\begin{tabular}{|l|l|l|l|}\hline
$X_0^0$ & $\tilde p_3$ & $\tilde p_2$ & $\tilde p_1$ \\\hline
& 24 & 24 & 24,30 \\\cline{3-4}
&    & 27 & 21,33 \\\cline{3-4}
&    & 30 & 24,30 \\\cline{2-4}
& 26 & 22 & 24,30 \\\cline{3-4}
&    & 28 & 18,36 \\\cline{3-4}
&    & 31 & 21,33 \\\cline{2-4}
& 28 & 23 & 21,33 \\\cline{3-4}
&    & 26 & 18,36 \\\cline{3-4}
&    & 32 & 24,30 \\\cline{2-4}
& 30 & 24 & 24,30 \\\cline{3-4}
&    & 27 & 21,33 \\\cline{3-4}
&    & 30 & 24,30 \\\hline
\end{tabular}
\hspace{2cm}b.\
\begin{tabular}{|l|l|l|l|}\hline
$X_1^{-1}$ & $\tilde p_3$ & $\tilde p_2$ & $\tilde p_1$ \\\hline
1. & 16 & 23 & 21,33 \\\cline{3-4}
   &    & 26 & 18,36 \\\cline{3-4}
   &    & 32 & 24,30 \\\cline{2-4}
   & 18 & 18 & 18,36 \\\cline{3-4}
   &    & 27 &  9,45 \\\cline{3-4}
   &    & 36 & 18,36 \\\cline{2-4}
   & 20 & 19 &  9,45 \\\cline{3-4}
   &    & 22 &  6,48 \\\cline{3-4}
   &    & 40 & 24,30 \\\hline
2. & 24 & 12 & 18,36 \\\cline{3-4}
   &    & 15 &  9,45 \\\cline{2-4}
   & 26 & 16 &  6,48 \\\cline{2-4}
   & 28 & 11 & 21,33 \\\cline{2-4}
   & 30 & 12 & 18,36 \\\cline{3-4}
   &    & 15 &  9,45 \\\hline
3. & 24 & 24 &     0 \\\cline{3-4}
   &    & 30 &     0 \\\cline{2-4}
   & 26 & 22 &     0 \\\cline{2-4}
   & 28 & 32 &     0 \\\cline{2-4}
   & 30 & 24 &     0 \\\cline{3-4}
   &    & 30 &     0 \\\hline
\end{tabular}

\renewcommand{\arraystretch}{1}

\vspace{.5cm}

\parbox{10.5cm}{
\noindent{\bf Table 4.}\ \ {\it Level 0 and 1 operators in the
total cohomology.}}
\end{center}

\vspace{.5cm}

The level 1 operators come in three branches. Solutions named 1. in
table 4.b have $x_1=x_2=0$, solutions 2. have $x_1=0$ and non-zero
$x_2,x_3$, and solutions
3. have non-zero $x_1,x_2,x_3$ but $p_1=0$.
Note that 1. and 2. are the level 1 $Q_1$ operators (table 3.b)
dressed to operators of the total
cohomology. The operators 3. correspond
to the level 0 $Q_1$ operators of
vanishing conformal weight (table 3.a).

Next we compute some screening charges, i.e. the charges that commute
with $Q$. First, we observe that all $Q_1$ screening currents
are $Q$ screening currents as well.
In particular, $S_x$ is still a screening current, and its associated
physical operator is again given by the relation (\ref{SxQ1}),
where now $P_3=[Q,\phi_3]$ has two extra terms
in addition to $[Q_1,\phi_3]$.
The physical operator $x^{-1}$, can be found using e.g.
the spectral sequence argument described at the
beginning of this section. We find that it is given
by (\ref{xinv1}) with the following modification:
\bea
x^{-1}\ \rightarrow\ x^{-1}-({\ts{15\over28}}c_2\partial c_4c_4
+{\ts{225\over2464}}q_2c_2c_3c_4)e^{ip_3\phi_3}\,,\ \
\tilde p_3=30\,.
\eea
In the total cohomology this is a level 4 operator.
Analogous physical operators $y,y^{-1},z,z^{-1}$ are also expected
to exist, where $y$ is supposed to have momentum $(0,-40,-20)$ and
$z$ should have nonzero $\phi_1$ momentum in order to connect states
with different $p_1$ values.

Four new screening currents involving $b_2$ and $\phi_1$ are
\bea
T_2^-&=&(1+{\ts{2\over3}}q_2b_2c_3)\Vp\,,
\ \ \ \tilde{\vec p}\equiv(\tilde p_1,\tilde p_2,\tilde p_3)=
(-12,12,0)\,,\nonumber\\
T_2^+&=&(1+{\ts{5\over6}}q_2b_2c_3)\Vp\,,
\ \ \ \tilde{\vec p}=(-15,15,0)\,,\nonumber\\
T_1^-&=&e^{ip_1\phi_1}\,,\ \ \ \ \tilde p_1=24\,,\nonumber\\
\label{T12}
T_1^+&=&e^{ip_1\phi_1}\,,\ \ \ \ \tilde p_1=30\,.
\eea
Now, remembering the $Q_1$ screening currents (\ref{T3m}) and (\ref{T3p}),
one can easily check that the
operators $T_i^{\pm}$, $i=1,2,3$, have exactly
the momenta of the standard screening
currents $S_i^{\pm}$ given by (\ref{CG}).
In fact, $T_1^{\pm}=S_1^{\pm}$ identically.
The other screening currents have been modified by a ghost contribution.
Probably, this is a consequence of the redefinition that we carried out
to obtain the $W_4$ BRST charge (\ref{j42}), since
in this redefinition the
scalar fields and ghosts are mixed to some extent \cite{redef,own3}.
An interesting observation was made in
\cite{DDR} where it was noted that the physical operators with standard
ghost structure, i.e. the level 0 operators, are precisely the composites
that can be formed out of the screening currents $S_i^{\pm}$.

Now that we have included the Virasoro constraint, it is trivial to obtain
screening currents associated to physical operators since the
descent equation (\ref{descent}) is solved by
$S_{\cal O}(w)=\oint {dz\over 2\pi i} b_2(z) {\cal O}(w)$.

In \cite{BMP2} a classification of physical states for a $W[g]$
minimal model coupled to $W[g]$ gravity is given.
These results have already been seen to agree with those of \cite{PopeC}
in the case of the two-scalar $W_3$ string (or pure $W_3$ gravity),
see refs. \cite{BMP2,BBRT}.

If we take $g=su(4)$
and the trivial ($c=0$) $W_4$ minimal model, we are able to compare
with our results.
Non-trivial cohomology classes exist, according to \cite{BMP2},
at the following values of the momentum
\bea
\label{BMP}
\vec p=w^{-1}(\alpha_-\sigma\vec\rho-\alpha_+\vec\rho)+
i\sqrt{2}\alpha_0\vec\rho\,,
\eea
where $\vec\rho$ and the parameters $\alpha_0,\alpha_{\pm}$ are
as before (see (\ref{alfa})), and $w$ is an element of the $su(4)$
Weyl group $W$ while $\sigma$ can be an element of the $\widehat{su(4)}$
affine Weyl group $\hat W$. The ghost number
at which the state with momentum
(\ref{BMP}) occurs is given by $-l_w(\sigma)$, where $l_w(\sigma)$ is
the twisted length of $\sigma$
\bea
\label{twl}
l_w(\sigma)=\lim_{N\rightarrow\infty}
(l(t_{-Nw\rho}\sigma)-l(t_{-Nw\rho}))\,,
\eea
and $l$ is the ordinary length of an affine Weyl group element.
In order to compute the twisted length, one should decompose the
translation
$t_{-Nw\rho}$ into the simple affine Weyl reflections
$\{\sigma_0,\sigma_1,\sigma_2,\sigma_3\}$ and then look for the
cancellations that take place between $t_{-Nw\rho}$ and $\sigma$.
For $\sigma\in W$, (\ref{twl}) reduces to $l_w(\sigma)=l(w^{-1}\sigma)
-l(w^{-1})$. The action of $\sigma_0$ on $\rho$ should be taken here as
$\sigma_0\rho=\sigma_{\theta}\rho+5\theta$, where $\theta$ is the
highest root of $su(4)$.
For completeness we give the decompositions of the translations
associated with the simple roots:
\bea
t_{e_1}&=&\sigma_2\sigma_3\sigma_0\sigma_3\sigma_2\sigma_1\,,
\nonumber\\
t_{e_2}&=&\sigma_3\sigma_1\sigma_0\sigma_1\sigma_3\sigma_2\,,
\nonumber\\
t_{e_3}&=&\sigma_2\sigma_1\sigma_0\sigma_1\sigma_2\sigma_3\,.
\label{st}
\eea

Taking $\sigma={\bf 1}$, (\ref{BMP}) yields all level 0 physical states
of table 4.a
when $w$ runs over the 24 elements of $W$.
Whereas the Weyl group action in (\ref{BMP}) does not change the level,
the affine Weyl group action does.
If we let $\sigma$ run over the
simple Weyl reflections $\{\sigma_1,\sigma_2,\sigma_3\}$
and $w$ over all elements in $W$, we obtain precisely all momenta
and ghost numbers of the level 1 prime physical states,
of which the ones with $G=-1$ have been listed in table 4.b.
So we find complete agreement with the results of \cite{BMP2}
on levels 0 and 1.
Also, the selection rule (\ref{SR}) is compatible with (\ref{BMP}).

The affine Weyl elements can be decomposed into ordinary Weyl
transformations and translations in the coroot lattice,
$\sigma=t_{\beta}w$. The translations associated with the
simple roots (\ref{st}) correspond to
the following changes in the momenta
\bea
\Delta_1 \tilde{\vec p}&=&(120,0,0)\,,\nonumber\\
\Delta_2 \tilde{\vec p}&=&(-60,60,0)\,,\nonumber\\
\Delta_3 \tilde{\vec p}&=&(0,-40,40)\,.
\eea
Physical operators with these momenta are supposed to be invertible
and they can be used to classify the complete cohomology
in terms
of a set of low level physical operators (see \cite{PopeC}).
Recall that
the operator $x$ with momentum $\tilde{\vec p}=(0,0,-30)$ can
also be used for this purpose. Using $x$ and the operators corresponding
to the simple root translations, an alternative basis
of $x$-like operators
is found to have momenta $(0,0,-30)\,,\ (0,-40,-20)$ and $(-60,-20,-10)$.

As before, we only considered one sector of the cohomology, the
prime operators of lowest ghost number.
There are seven other
sectors obtained by acting with the picture changing operators
$P_1,P_2\ {\rm and}\ P_3$.

\section{Discussion}
In this paper we have analysed the BRST cohomology of the
critical $W_4$ string.
Using the decomposition (\ref{j44}-\ref{j42}),
two minimal model structures have
been identified. The $Q_2$ cohomology
is seen to correspond to the $c={4\over5}$ $W_3$ minimal model.
There are
six primaries in this minimal model, with different highest
weights $(h,w)$. Moreover,
from the analysis it is clear that
the $Q_2$ cohomology contains an infinite
number of copies of this minimal model at different ghost numbers,
where operators with the same highest
weight $(h,w)$ can be connected by the
action of screening operators.

Since $Q_2$ only imposes the spin-four constraint, all $W_3$ descendants
of these highest weight operators are also $Q_2$ physical.
Moreover, these $W_3$ representations seem to be irreducible since the
null vectors are $Q_2$-exact (as far as we've checked it). This action
of $Q_2$ is reminiscent of that of a Felder BRST operator.

The $Q_1$ cohomology corresponds to the $c={7\over10}$ Virasoro
minimal model.
Again, all the primaries of the
Kac table have been found in the cohomology, and copies of these primaries
at different ghost numbers can be obtained by acting with screening
operators.
The descendants are also $Q_1$ physical apart from the null vectors
which seem to be $Q_1$-exact so that in this respect
$Q_1$ is like a Felder BRST operator
for this realisation of the $c={7\over10}$ Virasoro algebra.
The critical $W_4$ string shows clearly some resemblance to a non-critical
Virasoro string with $c={7\over10}$ minimal matter, and
somewhat less obvious it may resemble a non-critical $W_3$ string with
$c={4\over5}$ $W_3$ minimal matter.
Such relations were conjectured in \cite{own2}.

Finally, the cohomology of the total critical $W_4$ string contains
operators of zero total conformal dimension only, due to the Virasoro
constraint. The three-scalar critical $W_4$ string,
that we described here is alternatively called pure $W_4$ gravity
(as in e.g. \cite{DDR}). In any case, the spectrum coincides
(as far as we've been able to compare) with that
of $W_4$ gravity coupled to trivial $c=0$ matter \cite{BMP2}.
There are physical operators at ghost numbers ranging from minus infinity
to plus infinity, just as in the one-scalar Virasoro string and the
two-scalar $W_3$ string \cite{PopeC}. This seems to be typical for
minimal matter coupled to gravity \cite{LZ}.

The situation
for the more general non-critical $W_4$ string is much more complicated.
For instance, its BRST charge is not known explicitly,
although it may not be too difficult to extend the BRST charges of
\cite{W4BRST,own3} to the non-critical
form. Unfortunately, the redefinition
that simplifies the BRST charge can only be carried out in either the
matter or the gravity sector,
so that in the other sector the usual Miura realisation must be used.

It should not be difficult to generalize our results to the case of the
multi-scalar $W_4$ string. The \emt\ for $\phi_1$ can be replaced by
an arbitrary effective \emt\ $T_{eff}$ with the same central charge.
Thus multi-scalar
realisations of the $W_4$ symmetry can be obtained. Some three-scalar
states will then generalize to continuous momentum multi-scalar states,
some will generalize to discrete momentum multi-scalar states and some
may not be generalized at all to multi-scalar states.
An effective spacetime exponential
may be replaced by any effective spacetime
operator with the same OPE under $T_{eff}$ to
obtain other physical operators.
See \cite{PopeC}
for a discussion of this multi-scalar generalization
in the case of $W_3$.

As described in \cite{own3} it is expected that the
$W_N$ BRST charge can be decomposed in a way similar to the $W_4$ BRST
charge. This is certainly true at the classical level. The studies of
$W_3$ and $W_4$ strings make the following picture
of minimal models in the $W_N$ string very plausible.
Imposing the spin-$N$ constraint by its BRST operator $Q_N$ results
in an $(N,N+1)$ unitary $W_{N-1}$ minimal model. In the next step,
where the spin-$(N-1)$ constraint is added, the operators
are dressed to operators of the $(N,N+1)$ $W_{N-2}$ minimal model by the
BRST charge $Q_{N-1}$.
This goes on in the same way resulting in an $(N,N+1)$ Virasoro
minimal model in the $Q_1$ cohomology, and the total cohomology is
obtained from the double complex with BRST charges $Q_1$ and $Q_{Vir}$.
This agrees with the counting of central charges as
discussed in the introduction.
A similar discussion for non-critical $W_N$ strings may be found
in \cite{BBRT}.
Recent progress in constructing BRST charges for higher spin
strings (e.g. $W_{2,N}$ or $W_N$ strings) has been made in \cite{FWBRST}.

Some results on higher spin strings based on spin-$(2,N)$ $W$-algebras
have been obtained in \cite{PopeH1,PopeH2,PopeW24}.
A complication noted by the authors of those papers is that for $N\geq5$,
the central charge of the spin-$N$ sector, corresponding to
a $W_{N-1}$ minimal model, becomes greater or
equal to one, as can be seen also
from (\ref{cNN}). Consequently the number of effective
spacetime sectors which couple to the spin-$N$ fields
is no longer finite.
This complication is not expected to occur for the $W_N$ string, since
there is a sequence of $W_k$ minimal models of which the last one
is the $(N,N+1)$ Virasoro minimal model which of course has $c<1$
so that there is only a finite number of effective spacetime intercepts.

%W-constraints in matrix models for minimal matter coupled to 2d gravity

\section{Acknowledgements}
In most calculations we made use of the Mathematica package for computing
operator product expansions, OPEdefs, by K.~Thielemans \cite{OPE}.
I would like to thank Eric~Bergshoeff and Mees~de~Roo for discussions.
This work was performed as part of the research program of the
``Stichting voor Fundamenteel Onderzoek der Materie'' (FOM).

\end{document}